\documentclass[journal=jctcce,manuscript=article]{achemso}
\setkeys{acs}{usetitle = true}

\usepackage{chemformula} 
\usepackage[T1]{fontenc} 

\def\d{\delta}

\def\th{\theta}

\def\vf{\varphi}

\def\bra{\langle}
\def\ket{\rangle}

\def\xc{{\rm xc}}

\def\br{\mbox{\bf r}}
\def\bx{\mbox{\bf x}}

\newcommand*{\LightComments}{}%

\ifdefined\NoComments

\else
 \ifdefined\LightComments

 \else

 \fi
\fi
\author{Maria Hellgren}
\affiliation{Sorbonne Universit\'e, Mus\'eum National d'Histoire Naturelle, UMR CNRS 7590, Institut de Min\'eralogie, de Physique des Mat\'eriaux et de Cosmochimie (IMPMC), 4 place Jussieu, 75005 Paris, France}
\email{maria.hellgren@sorbonne-universite.fr}
\author{Tim Gould}
\affiliation{Qld Micro- and Nanotechnology Centre, Griffith University, Nathan, Qld 4111, Australia}
\title[Strong correlations in fractional KS theory]
  {Strong correlation and charge localisation in Kohn-Sham theories with \\fractional orbital occupations}
\abbreviations{IR,NMR,UV}
\begin{document}
\begin{abstract}
We study static correlation and delocalisation errors 
and show that even methods with good energies can yield
significant delocalization errors that affect the density, leading to large errors in predicting {\em e.g.} dipole moments. We illustrate this point by comparing
  existing state-of-art approaches with an accurate exchange 
  correlation functional based on a generalised valence-bond ansatz, 
  in which orbitals and fractional occupations are treated as 
  variational parameters via an optimized effective potential (OEP). 
  We show that the OEP exhibits step and peak features
  which, similar to the exact Kohn-Sham (KS) potential of DFT, 
  are crucial to prevent charge delocalization.
We further show that the step is missing in common
  approximations within reduced density matrix functional theory 
  resulting in delocalization errors comparable to those found in DFT approximations. 
  Finally, we
  explain the delocalization error as coming from an artificial mixing of the
  ground state with a charge-transfer excited state which is avoided if
  occupation numbers exhibit discontinuities.
\end{abstract}

\section{Introduction}
Electron delocalisation errors and related issues%
\cite{Yang2000,MoriSanchez2006,Cohen2008,MoriSanchez2009} have been
the subject of intensive study in recent years, due to their importance to molecular
modeling, particularly in charge transfer systems. Delocalisation errors arise in most common
\emph{ab initio} approximations;\cite{KohnSham,GGA,Becke1988} within
density functional theory (DFT) but also in more advanced many-body
methods such as MP2, $GW$ and
beyond.\cite{Bruneval2009,Hellgren2012-2,Caruso2013,Colonna2016} These errors are most prominent when breaking molecular bonds, and are responsible
for the dramatic failures to correctly predict the charge distribution at dissociation.\cite{Perdew1982,Makmal2011,Ruzsinszky2006}

In addition to delocalization errors much attention has been paid to overcoming the static correlation error in density functional approaches.\cite{Vydrov2007,Cohen2007,Baer2010,Tsuneda2010,%
  Johnson2011,Kraisler2013,Steinmann2013,Zheng2013,%
  Malet2012,Vuckovic2017}
This error is manifested in the strongly overestimated dissociation
energy of molecules composed of open-shell atoms. Very often static correlation and delocalisation errors occur simultaneously, exemplified by the LiH molecule. Consequently, such cases are
extremely difficult for most density functional approximations,
which reproduce neither correct densities nor energetics.
 
An alternative pathway is to use
reduced density matrix functional theory (RDMFT), by directly
approximating the first- and second-order reduced density
matrices.\cite{lowdinRDMFT,Donnelly1978,Yang2000}
The resulting energy expression then becomes a simultaneous
functional of (natural) orbital occupation factors, and a corresponding
set of orthonormal (natural) orbitals. With an appropriate choice of RDMFT
approximation, strong static correlation effects can then be directly
accommodated within a (natural) orbital framework.

It is also possible to accommodate orbitals within a DFT
framework, by employing an optimized effective potential (OEP)
approach\cite{OEP1,OEP2} together with an appropriate orbital
functional. In so doing, one hopes that an appropriate choice of
functional can lead to an approach that intrinsically avoids
delocalization errors. The best known OEP approach is the
exact-exchange (EXX) approximation within DFT. In EXX, one looks for
the Hartree-Fock energy of a system using orbital
functions $\vf_i$ satisfying a common, local Hamiltonian
\begin{align}
\hat h\vf_i(\br)=&[\hat{T}+V_{\text{EXX}}(\br)]\vf_i(\br)=\epsilon_i\vf_i(\br).
\label{eqn:KS}
\end{align}
EXX is known to reproduce many of the exact
features\cite{Schoenhammer1987,KLI1992,GoriGiorgi2009,%
Hellgren2012,Hellgren2013} of the true Kohn-Sham (KS)
potential and, as such, largely resolves the delocalization problem in single-reference systems.

However, unless spin symmetry is allowed to break, the EXX approximation does a poor job of predicting the energy and density of dissociation into open-shell atoms, where the outermost electrons are strongly correlated. The
majority of the error is caused by
a failure to localize the outermost electrons on the different odd
electron number atoms. Instead, the highest occupied molecular orbital
(HOMO, H) is doubly occupied, with electrons shared across the two
species. This allows ``ghost interactions'' to cause a static
correlation error, with energy cost
$1/2\bra\vf^2_{\rm H}|v|\vf^2_{\rm H}\ket$\cite{Hellgren2012-2}
between opposite-spin electrons that should be excluded by the
multi-determinant nature of the groundstate.

In this work we employ a multi-configurational approach that avoids ghost 
interactions and delocalisation errors via an optimized effective 
potential scheme. The energy is constructed from a wave function ansatz --
built from one-electron orbitals defined as non-orthogonal linear combinations of KS orbitals.
The KS orbitals are then optimized via an effective local potential.
This scheme differs from other OEP schemes utilising unoccupied
orbitals\cite{Gidopoulos2002,Grabowski2002,Hellgren2012-2} in
that the unoccupied orbitals are employed to remedy deficiencies
of the reference system - similar to the local RDMFT
approach \cite{lRDMFT} or 
other multi-configurational OEP methods.\cite{MCOEP1,MCOEP2}

We reveal the important role played by the effective potential
in dealing with charge localization, even in theories which depend
explicitly on occupation numbers, and how a failure to correctly
predict the potential (and thus orbitals) leads to delocalisation
errors even in methods that give, to leading order, correct
correlation energies. This error manifests in measurable
quantities, such as dipole moments.
\section{Optimized effective potential}
In this section we derive the OEP within theories that allow fractionally 
occupied KS orbitals. We employ a multiconfigurational approach similar to 
the ones proposed in Refs. \citenum{MCOEP1,MCOEP2} which we then, for 
two electrons, connect to the local RDMFT theory. \cite{lRDMFT}

\subsection{Wavefunction ansatz based on optimized Kohn-Sham orbitals}
The use of doubly occupied KS orbitals in DFT is an, in principle, 
exact construction. But, it is a bad starting point for treating
strongly correlated electrons which are spatially localised and thus
more naturally treated by localized orbitals.
Spin-DFT can sometimes achieve spatial localisation; at the expense
of breaking the fundamental spin symmetries. A proper
treatment of correlation is provided by multi-configurational
wavefunction approaches, but with the drawback that several 
configurations may be needed to converge the energy (or density).

In the following, we will explore the possibility of optimising the
orbitals via a KS system with a local potential, common to all orbitals. As a
proof-of-principle we propose a two-electron model that captures
challenging multi-reference physics, but is amenable to exact solutions
of all properties. We can thus use it to illustrate
important features of orbital-based approximations for multi-reference
physics, in a highly controlled fashion.

We construct a generalized valence-bond-like
ansatz\cite{GVB} of two Slater determinants built up from a set of nonorthogonal
correlated orbitals that are allowed to localise via the HOMO and LUMO
of the KS system
\begin{align}
  \psi_{A}(\br)=&\cos\theta\vf_{\rm H}(\br)+\sin\theta\vf_{\rm L}(\br)
  \label{quasio1}
  \\
  \psi_{B}(\br)=&\cos\theta\vf_{\rm H}(\br)-\sin\theta\vf_{\rm L}(\br).
  \label{quasio2}
\end{align}
Unlike the KS orbitals, $\vf_{\rm H}$ and $\vf_{\rm L}$,
the correlated orbitals are not eigenstates of
an effective Hamiltonian. Only at $\theta=0$ do both $\psi_A$ and
$\psi_B$ reduce to the KS HOMO orbital, and we recover double
occupancy. As $\theta$ is increased, more of the LUMO orbital is mixed
in and eventually at $\theta=\pi/4$, $\psi_A$ and $\psi_B$
become orthogonal. The more the KS orbitals are delocalized, the more
the correlated orbitals will be localized, and vice versa. 

The two orbitals are then used to construct a  singlet wavefunction
\begin{align}
|\Psi \ket=\frac{1}{\sqrt{N}}
\{|A\!\!\uparrow\!B\!\!\downarrow\ket
+ |B\!\!\uparrow\!A\!\!\downarrow\ket\}
\label{ansatz}
\end{align}
where $N=4(\cos^4\th+\sin^4\th)$ is the normalization constant. 
We notice that when $\theta=0$ this is just the Hartree-Fock (HF) 
wave function ansatz. We can also write 
\eqref{ansatz} as
\begin{align}
|\Psi \ket=\frac{1}{\sqrt{N}}[\psi_{A}(\br)\psi_{B}(\br')+\psi_{B}(\br)\psi_{A}(\br')]|S \ket
\label{abwave}
\end{align}
where $|S \ket$ is the spin-singlet wavefunction.

Taking the expectation value of the full many-body Hamiltonian with
respect to the wave function in Eq.~\eqref{ansatz} we find the density 
and kinetic energy to be
\begin{align}
  &n^{\th}(\br)= 2f_{\rm H}|\vf_{\rm H}(\br)|^2+2f_{\rm L}|\vf_{\rm L}(\br)|^2
  \equiv \bra\Psi|{\hat{n}(\br)}|\Psi\ket
  \label{dens}
\end{align}
\begin{align}
  &T_s^{\th}= 2f_{\rm H}\bra\vf_{\rm H}|\hat{T}|\vf_{\rm H}\ket
  + 2f_{\rm L}\bra\vf_{\rm L}|\hat{T}|\vf_{\rm L}\ket
  \equiv \bra\Psi|\hat{T}|\Psi\ket.
  \label{TJ}
\end{align}
The occupation numbers $f_H=4\cos^4\th/N$ and $f_L=4\sin^4\th/N$
are related to the fourth power of the expansion coefficients of the 
correlated orbitals $\psi_A$ and $\psi_B$ and the factor of two comes from 
the spin-summation. 

The electron-electron interaction is given by
\begin{align}
  E_{\rm Hxc}^{\th} =&\frac{2}{N}\int \!d\br d\br'\,
  |\psi_A(\br)|^2v(\br,\br')|\psi_B(\br')|^2 \nonumber\\
  &+\frac{2}{N}\int\! d\br d\br'\,
  \psi_A(\br)\psi_B(\br)v(\br,\br')
  \psi_A(\br')\psi_B(\br')\nonumber\\
  =& \int\! d\br d\br'\, n_{\rm HL}^{1/2}(\br,\br') v(\br,\br')
  n_{\rm HL}^{1/2}(\br',\br)
  \label{2el}
\end{align}
where
\begin{align}
n^{1/2}_{\rm HL}(\br,\br')=\sqrt{f_{\rm H}}\vf_{\rm H}(\br)\vf_{\rm H}(\br')-\sqrt{f_{\rm L}}\vf_{\rm L}(\br)\vf_{\rm L}(\br').
\label{densmathl}
\end{align}
We have called this term the Hartree and exchange and correlation (Hxc) energy but we note that it does not contain kinetic correlation contributions.

Finally, we use Eqs. \eqref{dens}-\eqref{2el} to generate the total energy 
\begin{align}
  E=\min_{\th}\bigg[
  T_s^{\th}+\int \!d\br \,V_{\rm Ext}(\br)n^{\th}(\br)+E_{\rm Hxc}^{\th}
  \bigg],
  \label{eqn:QEXXth}
\end{align}
where the energy minimization over $\theta$ determines the amount of
localization of $\psi_{A/B}$, via its optimizing value $\theta_0$.
We henceforth implicitly define quantities using $\theta=\theta_0$ and
drop the superscript $\theta$, e.g. $n\equiv n^{\theta_0}$.

The ansatz described above is known to capture H$_2$ dissociation 
rather well if the orbitals are expressed in a Gaussian basis 
and coefficients are optimized variationally. The crucial difference 
in the approach presented here is that the orbitals 
are expressed in terms of the HOMO/LUMO KS orbitals that are optimised via an 
effective local KS potential. 
This potential is derived from the condition that the total energy at every fixed value of $\theta$ is stationary with respect to variations of the total potential $V_{s}(\br)=V_{\rm Ext}(\br)+V_{\rm Hxc}(\br)$, i.e.,
\begin{align}
  \frac{\delta E^{\th}}{\delta V_s}=
  \frac{\delta T_s^{\th}}{\delta V_s}+\int \! \,V_{\rm Ext}\frac{\delta n^{\th}}{\delta V_s}+\frac{\delta E_{\rm Hxc}^{\th}}{\delta V_s}=0.
  \label{dedv0}
\end{align}
This leads to an OEP type of equation for the Hxc potential
\begin{align}
\int \!d\br' \frac{\delta n^{\theta}(\br')}{\delta V_{s}(\br)}
V_{\rm Hxc}(\br')=\frac{\delta E^{\theta}_{\rm Hxc}}{\delta V_{s}(\br)}.
\label{oep}
\end{align}
The variation of the density with respect to $V_s$ can be obtained from first order (static) perturbation theory 
\begin{align}
\frac{\delta n^{\th}(\br')}{\delta V_{s}(\br)}=\sum_{i=H,L}\sum_{j\ne i}4f_i\frac{\varphi_i(\br)\varphi_j(\br)\varphi_i(\br')\varphi_j(\br')}{\varepsilon_i-\varepsilon_j}\;.
\label{resp}
\end{align}
The variation of the Hxc energy is straightforwardly
evaluated using Eq. (\ref{2el}), giving
\begin{align}
\frac{\delta E^\th_{\rm Hxc}}{\delta V_{s}(\br)}=&\sum_{i=H,L}\sum_{j\ne i}4f_i\frac{\varphi_i(\br)\varphi_j(\br)}{\varepsilon_i-\varepsilon_j}\int\! d\br'' d\br'\, \varphi_i(\br'')\varphi_j(\br'') v(\br'',\br')|\varphi_i(\br')|^2 \nonumber\\
&-\sum_{j\ne H}4\sqrt{f_H}\sqrt{f_L}\frac{\varphi_H(\br)\varphi_j(\br)}{\varepsilon_H-\varepsilon_j}\int\! d\br'' d\br'\, \varphi_H(\br'')\varphi_j(\br') v(\br'',\br')\varphi_L(\br')\varphi_L(\br'') \nonumber\\
&-\sum_{j\ne L}4\sqrt{f_H}\sqrt{f_L}\frac{\varphi_L(\br)\varphi_j(\br)}{\varepsilon_L-\varepsilon_j}\int\! d\br'' d\br'\, \varphi_L(\br'')\varphi_j(\br') v(\br'',\br')\varphi_H(\br')\varphi_H(\br'').
\end{align}

The resulting OEP equation is very similar in structure to the exact-exchange OEP equation. Therefore, similar numerical techniques can be used to solve it. In this work we have expanded both orbitals and orbital products in a spline basis set. In this way, the OEP equation becomes a matrix equation, in which the potential is obtained by inverting the response matrix [Eq. (\ref{resp})] in a subspace defined by fixing the HOMO energy to, e.g., the ionization energy. 
For more details regarding the OEP implementation see Refs. \citenum{hellgren07,hellgren09}. 

The main difference to previous work is that, here, the potential is
evaluated at some starting value $\theta_1$ for which the corresponding
total energy and orbitals are calculated. This procedure is repeated
until the minimum energy at $\theta_0$ is found. Testing revealed
that the two minimisations (over $\theta$ and the orbitals) can be
carried out in either order. We optimise the orbitals first, and
then the angle.

We note that the effective Hxc potential in Eq.~(\ref{oep}) is
different from the standard DFT Hxc potential with integer occupation
numbers but serves a similar role. It is, in fact, equivalent to a
potential generated through excited-state ensemble DFT\cite{GOK1,GOK2}
and it thus inherits the usual good properties of KS potentials, such as
uniqueness; and is fully amenable to the usual KS mechanics, such as
division into kinetic, Hartree-exchange and correlation
components.\cite{Gould2017-Limits,Gould2019-DD} Note, however, that within this formalism
the division of Hartree, exchange and correlation terms into kinetic and
electrostatic components differs slightly from conventional DFT. Their
sum, nonetheless, remains the same.

Although we restrict ourselves to two-electron systems the combination
of a wavefunction ansatz with orbitals optimised via a KS potential can
be made more general.\cite{MCOEP2} In
this work our aim is to study the reliability and exact features of
this local KS potential (from now on referred to as COEP
[correlated orbital OEP]) in situations where density errors can be
significant. We thus seek to reveal features of the exact potential that must
be dealt with by approximations for more complex cases.

\subsection{Connection to reduced density matrix functional theory}
The reduced density matrix of the many-electron wavefunction is defined as
\begin{align}
n(\bx,\bx')=N\!\int\!d\bx_{2-N}\Psi(\bx,\bx_2\ldots\bx_N)\Psi^*(\bx',\bx_2\ldots\bx_N)
\end{align}
where the variable $\bx$ carries both space and spin degrees of freedom. It is often written in terms of its spin eigenfunctions (natural orbitals) $\phi_i(\bx)$ and eigenvalues (occupation numbers) $f_i$
\begin{align}
n(\bx,\bx')=\sum_i f_i\phi_i(\bx)\phi^*_i(\bx').
\end{align}
It can be directly shown that the natural orbitals of the wavefunction in
Eq.~\eqref{abwave} are just the KS 
HOMO/LUMO orbitals, and the occupation numbers are given by the factors $f_{\rm H/L}$ 
defined above. It is then easy to see that the derivation leading to the expression 
for the total energy Eqs. (\ref{TJ}-\ref{2el}) is valid also for the exact wavefunction 
provided it is expressed in terms of its natural orbitals and occupation numbers.\cite{lowdinRDMFT}
The COEP approximation defined in the previous section can thus alternatively be derived from 
the exact RDMFT energy expression with the restriction of keeping only two natural orbitals, further assumed to be generated by an effective local potential. The use of a local potential in RDMFT is always approximate 
but recent studies suggest that it does not impose a severe restriction. \cite{lRDMFT} For this reason we refer to COEP as a ``nearly exact'' model, in the sense that it
accurately captures strong correlations, and misses only a small fraction
of dynamic correlations due to the restriction of keeping only two orbitals.

A common approximation to the exchange correlation energy in RDMFT is the 
M\"uller functional\cite{muller,bb} obtained by taking the square root of the occupation 
numbers in the Fock exchange term. It can thus be written as
\begin{align}
E_{\xc}^{{\rm M\ddot uller}}=-\frac{1}{2}\int\! d\bx d\bx' n^{1/2}(\bx,\bx') v(\br,\br')
n^{1/2}(\bx',\bx)
\label{muller}
\end{align}
where $n^{1/2}(\bx,\bx')=\sum_i\sqrt{f_{i}}\phi_{i}(\bx)\phi_i(\bx')$. The corresponding local M\"uller potential at fixed occupations is determined by an OEP equation like Eq. (\ref{oep}) but with the xc part of the energy given by Eq. (\ref{muller}), and the Hartree part calculated in the
conventional way.\cite{lRDMFT} 

With some  similarities to the nearly exact functional
[Eq. (\ref{2el})] the M\"uller functional  captures static
correlation rather well but tends to overestimate the correlation 
energy. In Ref. \citenum{hubbardRDMFT} it was shown that spurious
magnetic instabilities  occur in the Hubbard dimer and that an
energy discontinuity is missing. Although simple 
improvements such as changing the square root of the occupation
numbers to an optimised power improve energies\cite{powerRDMFT}
fundamental issues remain.\cite{hubbardRDMFT}

In the next section we will study how well the M\"uller functional
performs in  dissociating a heteronuclear dimer. We restrict
the functional to HOMO/LUMO only, in order to maximize its similarity
to COEP defined above. We will thus show that despite energies of
similar quality to COEP, the delocalisation error remains as large as within a typical DFT approximation. We will 
further demonstrate that this error is related to a missing step
of the M\"uller effective potential that is correctly captured by
the COEP effective potential.

\begin{figure*}[t]
  \includegraphics[width=0.45\linewidth]{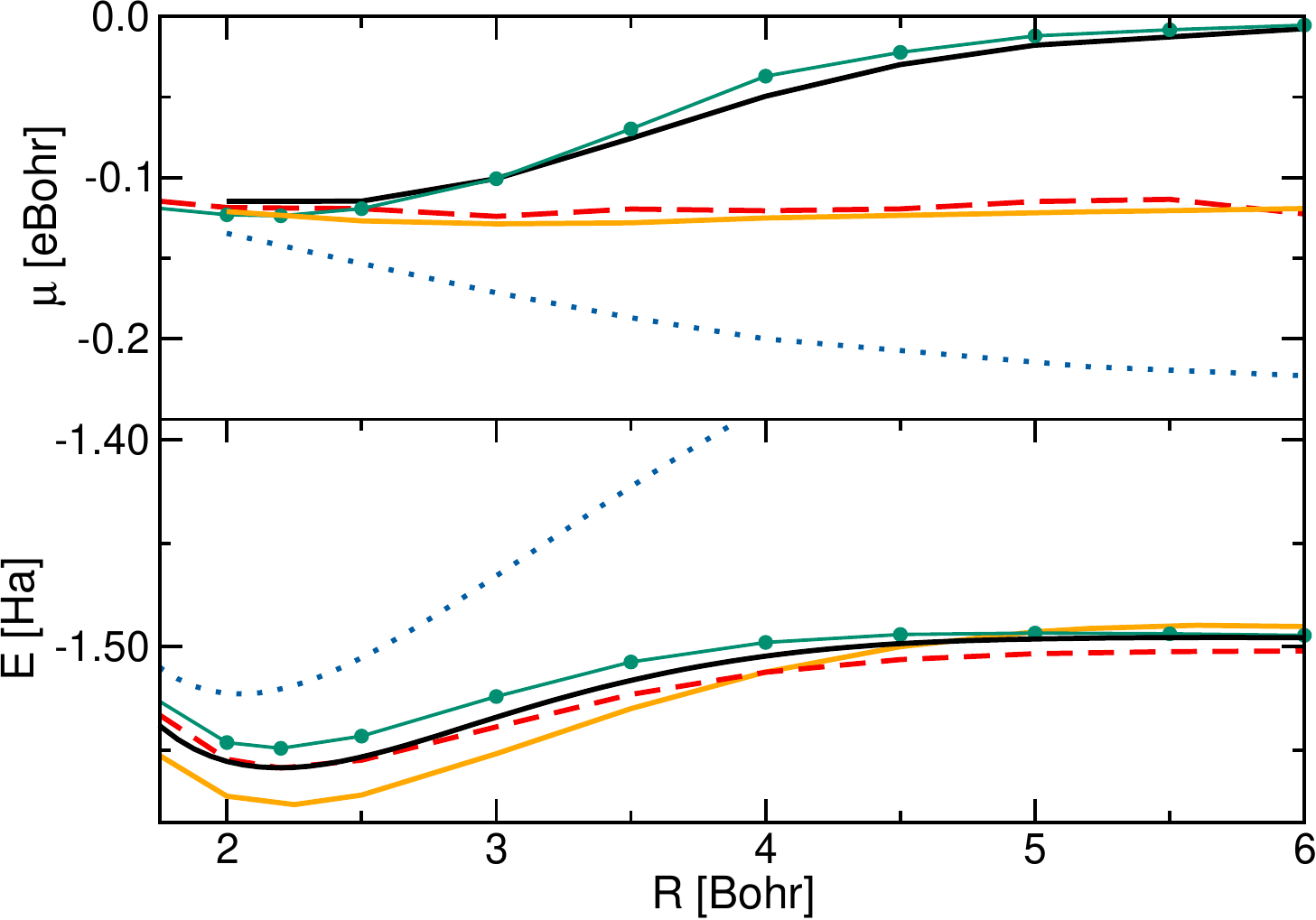}\hspace{10mm}
  \includegraphics[width=0.45\linewidth]{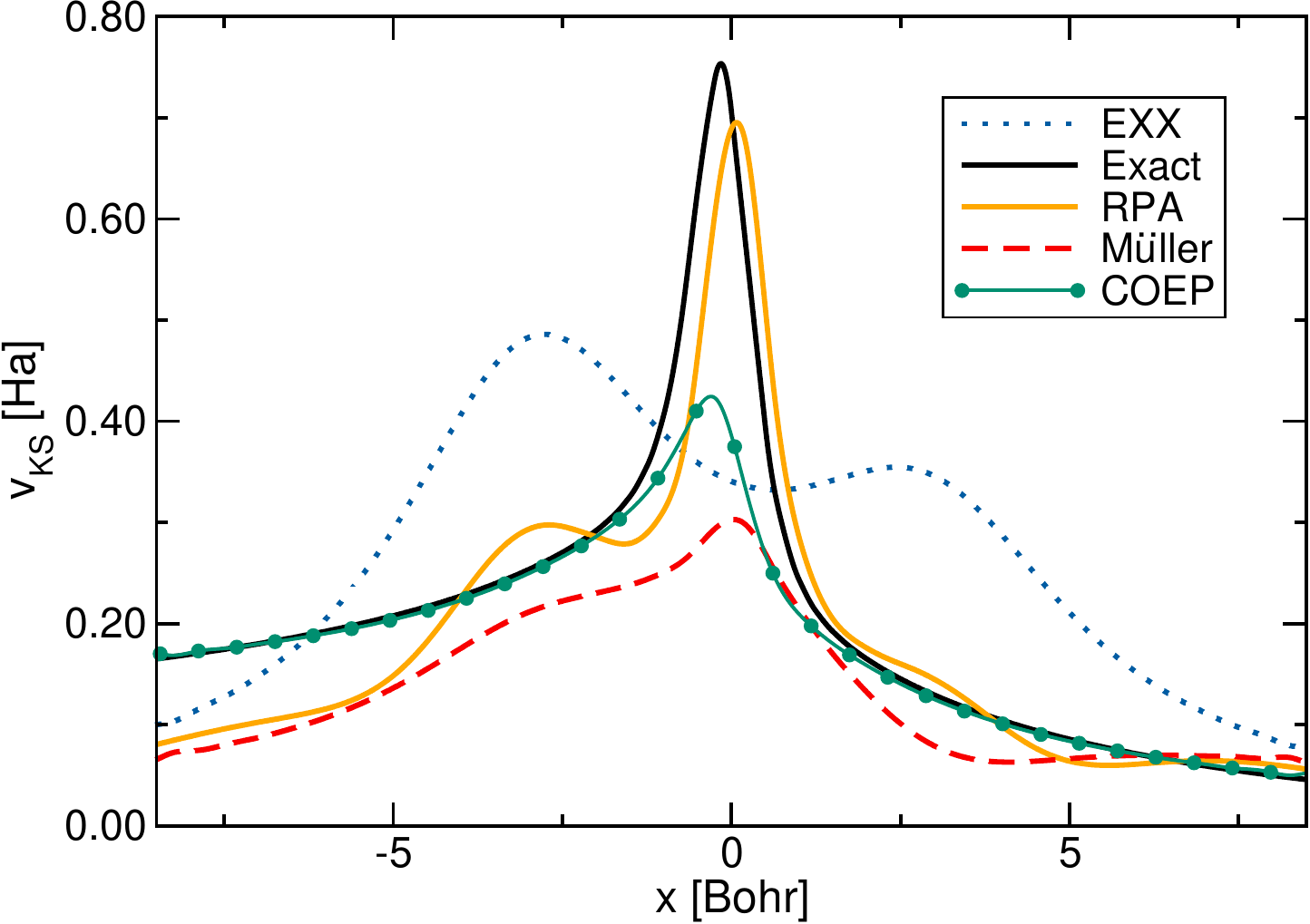}
    \caption{Left: The total energy $E$ and dipole moment $\mu$
    (which highlights errors in the density) of the
      LiH molecule with EXX (blue dots), RPA (orange full), M\"uller
      (red dashed) and COEP (green filled circles) compared to exact
      results (black full). Right: The corresponding KS potentials at
      separation $R=6$ Bohr.
    \label{lihendifig}}
\end{figure*}

\section{Numerical analysis}
In this section we present a numerical analysis of a stretched heteronuclear dimer that mimics LiH-like molecules at dissociation. To this end, we introduce soft-coulomb interactions in one dimension.\cite{Maitra2009,helbig_soft,Oueis2018} Thus, the electron-electron interaction is $v(x-x')=1/\sqrt{1+(x-x')^2}$ and the external potential is
\begin{align}
V_{\rm Ext}=&\frac{Z_1}{\sqrt{1+(x+R/2)^2}}+\frac{Z_2}{\sqrt{1+(x-R/2)^2}},
\label{eqn:VExtH}
\end{align}
where $Z_1=-1.2$ and $Z_2=-1$. The difference between the ionization energies of the two 1D atoms is 0.2Ha, which can be compared to the LiH value of 0.3Ha. Other choices of parameters are also possible.\cite{Maitra2009,Oueis2018} The core electrons on the 1D `Li' atom are frozen and hence 
the `LiH' is a two electron system for which we have a numerically exact solution to
compare with. Calculations are done in a 1D spline basis-set previously
proven  suitable to solve OEP-type of problems.\cite{hellgren07,hellgren09} 
LDA and exact results are obtained with the {\sc OCTOPUS} code.\cite{helbig_soft,octopus} 

All calculations are performed self-consistently.
The potential for the COEP and M\"uller functionals are defined by applying Eq.~\eqref{oep} to
Eq.~\eqref{eqn:QEXXth} and Eq.~(\ref{muller}), respectively. The random-phase approximation (RPA) potential is found via the linearized Sham-Schl\"uter equation as described in previous works.\cite{lss,casida,vonBarth05,hellgren07}

We will start by looking at the potential energy curve $E$ and dipole
moment $\mu$ and then analyse the ability of approximate functionals to deal with charge localisation, and relate this ability to energy derivative discontinuities and step features in the Hxc potential.
\subsection{LiH dissociation}
In a stretched symmetric or homonuclear system like H$_2$ the KS HOMO and LUMO are 
both delocalized by symmetry contraints. As a consequence, 
when $\theta=\pi/4$ the correlated orbitals are completely localized, and, due to the
infinitesimal overlap between $\psi_A$ and $\psi_B$,
the interaction energy in Eq.~\eqref{2el} is close to zero. 
There is thus no static correlation error in COEP.

In the more challenging heteronuclear systems the KS
orbitals may, or may not, be delocalized, depending crucially on the
behavior of the KS potential. If the electrons were completely non-interacting
the HOMO/LUMO would be localized on the atom with stronger/weaker nuclear 
potential. Hence, in the dissociation limit we would find both electrons on the 
atom with the largest ionization energy and zero on the other. 
In the LiH molecule the repulsive electron-electron interaction changes the 
charge distribution favouring one electron on each atom in the
dissociation limit. This system thus provide a critical
test of the quality of electron densities obtained via
approximations, quantified through the quality of the dipole moment
$\mu=\int [n(x)-n_{\text{nuclear}}(x)] x dx$.

Standard KS DFT, which relies on a doubly occupied HOMO, solves this problem 
by aligning the effective HOMO of the two atoms via a step in the KS
potential when calculated exactly.\cite{Perdew1982,maitraCT} 
In this way the orbitals delocalize and the doubly occupied HOMO 
simulates localized electrons. However, all known approximate
functionals in DFT
fail to achieve the step when open shell atoms are dissociated.
Consequently, fractionally charged atoms are found in the dissociation
limit together with a diverging dipole moment: $|\mu(R)|\to\infty$
for $R\to\infty$.

How functionals that exploit fractionally occupied KS orbitals deal with
the same situation is presently not well understood. From our accurate COEP 
approximation (Eq.~\eqref{2el}) we see that the KS orbitals \emph{must}
delocalize to ensure the interaction 
energy vanishes in the dissociation limit. Therefore, we expect a step
feature similar to the one in  DFT to develop in the KS potential of
Eq.~\eqref{oep}.

In Fig. \ref{lihendifig} (left) we show the potential energy curves 
and dipole moment of our model LiH molecule using the M\"uller, RPA, EXX 
and COEP functionals and compare them to the exact result. 
As for H$_2$, we see that both M\"uller and COEP 
are close to free from static correlation errors and have
similar overall performance on energies. M\"uller out-performs COEP near contact due to a better treatment of dynamical correlation. But it becomes worse during dissociation.
RPA also performs well near contact and in the dissociation limit. Nonetheless,
all approximations (except EXX) give satisfactory dissociation
curves and total energies.

However, we see a drastically different behaviour for the
dipole moment, $\mu$. The dipole moment serves as an indirect
measure of the charge distribution, as discussed above.
It has previously been used to assess the quality of densities from
quantum chemical approximations\cite{Hait2018}.
While COEP gives very accurate results, both M\"uller and RPA fail.
A diverging dipole moment implies that the molecule dissociates into 
fractionally charged atoms. This problem, previously discussed for
RPA,\cite{Hellgren2012-2,MORI-RPA,Colonna2016} has so far not been
recognized for functionals in RDMFT. Our work thus establishes
that RDMFT is not a guaranteed solution for this problem as it
is not free from the delocalization errors that plague DFT.

\begin{figure*}[t]
  \includegraphics[width=0.97\linewidth]{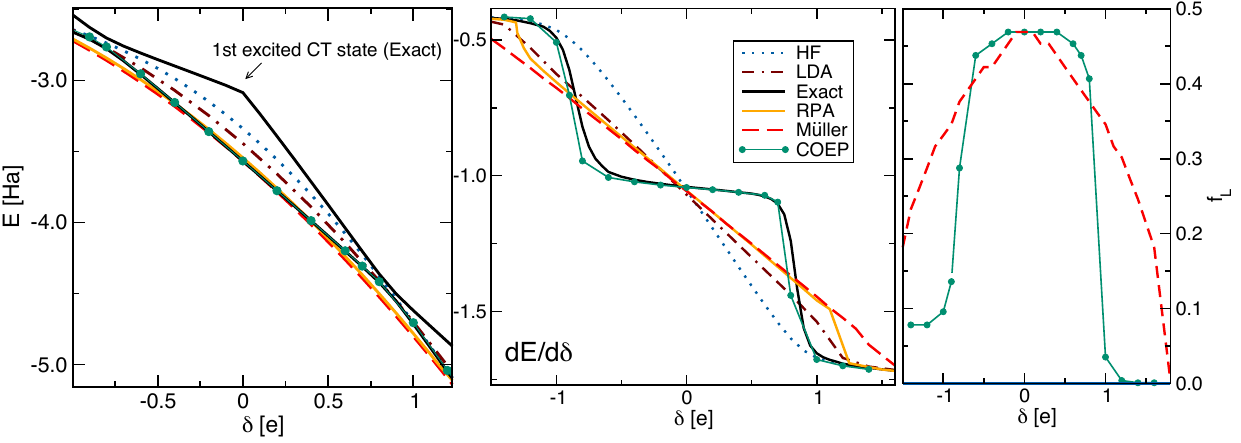}
  \caption{Left: Total energy as a function of the strength of the
    potential of atom 2 ($Z_2=-2-\d$) with HF, LDA, RPA, M\"uller and
    COEP compared to exact result. Also the first excited singlet
    state is shown. Middle: Derivative of the ground state energy with
    respect to $\d$ for the different approximations. Right: The
    corresponding occupation of the LUMO orbital within M\"uller and
    COEP.}
  \label{fracfig}
\end{figure*}

\subsection{The role of the potential}
By looking at the corresponding KS potentials (Fig.~\ref{lihendifig}, right)
the origin of the density error becomes clear. The exact KS potential of DFT 
(black full line) exhibits a peak at around the bond midpoint,
and a step that aligns the ionization potentials of the atoms.
The peak has its origin in kinetic correlation and acts to separate
the charges.\cite{peak1,peak2} This peak is present also in the RPA
but cannot alone prevent
delocalisation.\cite{Hellgren2012-2,Colonna2016}
Due to the improved description of kinetic correlation in KS theories
with fractional orbital occupations, like RDMFT, the peak is expected
to be less relevant. This is, indeed, the case as demonstrated by the
much smaller peak found in COEP.

We next turn our attention to the step. The step feature can be observed already in the potential of the EXX approximation when dissociating molecules into closed shell fragments\cite{Hellgren2012} or when applying an electric field to molecular chains.\cite{gisbergen99,kummel04,Hellgren2013} However, in the case of LiH-like dissociation none of EXX, RPA or the
M\"uller approach exhibit this feature. In contrast, COEP exhibits a
step of exactly the same size as the exact KS potential of DFT. The
missing step in both M\"uller and RPA explains the error in the
their densities; and hence the divergent dipole moment.

From this simple demonstration we can conclude that the step is not
only an essential feature of the DFT KS potential but also of the
local potential used in local RDMFT theory or in wavefunction/ensemble
theories that exploit optimized KS orbitals.
We emphasise that orbital optimisation in LiH-like molecules is crucial as 
the EXX orbitals are clearly unsuitable for either accurate energies
or densities.

Finally, we note that our work closely parallels recent results from
Oueis and Wassermann\cite{Oueis2018} in the context of \emph{exact}
partitioning theories. They studied the exact pinning potential of
dissociated 1D H$_2$ and LiH (using similar electronic models to those considered here).
The pinning potential, which incorporates non-local quantum effects to
supplement local models based on partitioning of the diatoms into atoms,
was shown to have analogous step and peak features to those found here,
which are caused by the same fundamental physics.

\subsection{Delocalisation error and derivative discontinuity}
We will now analyse in more detail the origin of the step in the KS
potential. In DFT the step has been related to derivative
discontinuities of the exact xc energy functional when studied as a function of
particle number.\cite{Perdew1982,stepshodgson} To expose these
discontinuities an extension of DFT functionals to ensembles composed
of states with different particle numbers is required. Most
approximate functionals have, however, been shown to miss the derivative
discontinuity leading to an incorrect dissociation limit with atomic
densities integrating to a fractional number of particles. 
The delocalisation error is therefore sometimes called fractional charge error.\cite{MoriSanchez2009}

To continue our analysis, and uncover the properties of functionals
involving fractionally occupied KS orbitals we will use an 
alternative approach, introduced in Ref. \citenum{fracnuclei},
which does not involve a direct treatment of
ensembles. Specifically, we study a stretched molecule at large
separation and vary the nuclear charge on one of the atoms to
transform the properties of the groundstate system. In our
model system we set $R=5$, set the charge on the first atom to
$Z_1=-2$ and, set the second to $Z_2=-2-\d$ where $\d$ is varied
between -1.5 and 1.5. We then calculate the total ground state
energy as a function $\d$. By varying $\d$ we are able to
change the number of electrons that localize on the two atoms.
For certain values of $\d$ (e.g. $\d\approx-0.9$), the number or
left and right electrons can be varied continuously and the two atoms
attain fractional charges, as detailed below.
Since the well-separated atoms are only weakly coupled,
they become closely analogous to \emph{isolated}
atoms attaining fractional charges via interactions with a bath.

In the exact calculation we have access to both ground and excited
states. When $\d~<-0.9$ the exact ground state  $|\Psi_0\ket$ has two
electrons on atom 2 and none on atom 1. The first {\em singlet}
excited state $|\Psi_{1S}\ket$ corresponds to a charge transfer state
where one electron on atom 2 is transferred to atom 1
(see, e.g., Refs. \citenum{Maitra2009,Gould2018-CT}). At around
$\d=-0.9$, $|\Psi_0\ket$ and $|\Psi_{1S}\ket$ become nearly degenerate
(exactly so in the limit of infinite separation) and it is possible 
to form an arbitrary linear combination of them that has (almost) the
same energy. These states are allowed to have fractional charges on
the atoms even at infinite separation and hence the situation corresponds to the one of studying one atom in an ensemble of states. As
soon as $\d>-0.9$, the excited state with one electron on each atom
(i.e. the first excited charge-transfer state) becomes the ground state and vice versa.

In Fig. \ref{fracfig} we plot the energy and its derivative with
respect to $\d$ as functions of $\d$ for the exact case, COEP,
M\"uller, EXX, LDA and RPA. In the energy plot we also include the energy
of the first exact singlet excited state. The most
notable feature is that, in the exact calculation, we see kinks or
derivative discontinuities appearing at the "switching points" between
the ground and the excited state. These discontinuities becomes
sharper the further apart the atoms are.

Turning to the approximations, we see that EXX performs well until $\d$ crosses 
the point where the ground- and excited state become degenerate in the exact calculation. 
As soon as the charge-transfer state should be favoured the EXX predicts a 
mixture of the two states leading to spurious fractional charges on the atoms. This error is consistent with the large
delocalisation errors found at dissociation of LiH (see
Fig. \ref{lihendifig}) and allows us to view the delocalisation (or fractional charge) error as an artificial mixing of
the ground state with the first excited charge-transfer state beyond
the point of degeneracy. A similar interpretation is often given to
the fractional spin error but as an artificial mixing with the first
{\em triplet} excited state (not shown in the figure as spin-symmetry
is restricted here). 

The connection between the derivative discontinuity in the energy and the step in potential has been made previously in the literature.\cite{Perdew1982,stepshodgson} In addition, it has been demonstrated that the development of the potential step during dissociation is connected to the appearance of an avoided crossing between the ground state and a charge-transfer excited state.\cite{Maitra2009} Our analysis shows how these two pictures can be merged. We also note that when dissociating a strongly ionic bond as in NaCl the avoided crossing could become close to infinitely sharp,\cite{Ruzsinszky2006} a situation more similar to the "switching point" in our model at large separation but, in this case, occurring already at finite separation.  

The RPA clearly performs better than EXX. Nonetheless, the derivative
discontinuities are missed, leading to delocalisation errors of
similar magnitude to LDA.\cite{Colonna2016} In view of the
results in the previous section it is not surprising to see that the
M\"uller functional fails as badly as RPA. Despite having an
ostensibly good treatment of strong correlations, it lacks the ability
to correctly treat charge localization.

Only the COEP, which is close to the exact functional in RDMFT, is able to reproduce the correct behaviour
with an energy discontinuity, or a step in the derivative. We note
that restricting to a local potential does not impose any severe
restriction even in this challenging situation. This can thus not be
blamed for the failure of the M\"uller functional, for example.

In the right panel of Fig. \ref{fracfig} we plot the
occupation numbers in COEP and the M\"uller functional.  While the
occupation numbers of the M\"uller functional are smooth as a function
of $\d$, the occupation numbers of COEP exhibit discontinuities at the
points where the energy has derivative discontinuities. Thus, just
as in DFT, the problem of delocalization in RDMFT can be related to
missing discontinuities in the energy functional. Fig. \ref{fracfig}
(right) provides evidence that these discontinuities appear directly
in the occupation numbers whenever charge localization is 
appropriately dealt with, as in COEP.
 
\section{Conclusions}
In this work we have analyzed the optimized effective potential within approaches that deal explicitly with fractional occupation numbers. In particular, we have studied the challenging case of heteronuclear dissociation and compared the potential to the KS potential of DFT, showing that, e.g., the step remains an essential feature. We have also introduced a generally applicable stringent test that highlights the ability of approximate functionals in dealing with charge localization.

We showed that RDMFT approximations such as the M\"uller functional, just like RPA,  miss an important energy derivative discontinuity, and thus suffer from rather large
delocalization errors that lead to errors in the
dipole moment. Furthermore, by employing
a multiplicative OEP in the nearly exact COEP, we showed that these errors are avoided thanks to the appearance of the step features in the
effective potential.

We thus recommend further attention be placed on the quality of the potential in theories that exploit fractional
orbital occupation numbers such as RDMFT and, more generally, in wavefunction or ensemble
theories. Our work indicates that using optimized effective potentials
can significanly improve on existing techniques, and lead to improvements in the
current state-of-art in electronic structure theory.

Finally, we note that these results have some implications
in the context of density-driven and functional-driven
analysis\cite{Kim2013} of RDMFT. The M\"uller functional gives good
energies of LiH dissociation, despite rather poor densities,
implying small density-driven errors.
This behaviour should be explored in future work.
\begin{acknowledgement}
M.H. acknowledges Dr. Florian G. Eich for valuable discussions on RDMFT.
\end{acknowledgement}


\begin{mcitethebibliography}{68}
\providecommand*\natexlab[1]{#1}
\providecommand*\mciteSetBstSublistMode[1]{}
\providecommand*\mciteSetBstMaxWidthForm[2]{}
\providecommand*\mciteBstWouldAddEndPuncttrue
  {\def\EndOfBibitem{\unskip.}}
\providecommand*\mciteBstWouldAddEndPunctfalse
  {\let\EndOfBibitem\relax}
\providecommand*\mciteSetBstMidEndSepPunct[3]{}
\providecommand*\mciteSetBstSublistLabelBeginEnd[3]{}
\providecommand*\EndOfBibitem{}
\mciteSetBstSublistMode{f}
\mciteSetBstMaxWidthForm{subitem}{(\alph{mcitesubitemcount})}
\mciteSetBstSublistLabelBeginEnd
  {\mcitemaxwidthsubitemform\space}
  {\relax}
  {\relax}

\bibitem[Yang \latin{et~al.}(2000)Yang, Zhang, and Ayers]{Yang2000}
Yang,~W.; Zhang,~Y.; Ayers,~P.~W. Degenerate Ground States and a Fractional
  Number of Electrons in Density and Reduced Density Matrix Functional Theory.
  \emph{Phys. Rev. Lett.} \textbf{2000}, \emph{84}, 5172--5175\relax
\mciteBstWouldAddEndPuncttrue
\mciteSetBstMidEndSepPunct{\mcitedefaultmidpunct}
{\mcitedefaultendpunct}{\mcitedefaultseppunct}\relax
\EndOfBibitem
\bibitem[Mori-S\'{a}nchez \latin{et~al.}(2006)Mori-S\'{a}nchez, Cohen, and
  Yang]{MoriSanchez2006}
Mori-S\'{a}nchez,~P.; Cohen,~A.~J.; Yang,~W. Many-electron self-interaction
  error in approximate density functionals. \emph{J. Chem. Phys.}
  \textbf{2006}, \emph{125}, 201102\relax
\mciteBstWouldAddEndPuncttrue
\mciteSetBstMidEndSepPunct{\mcitedefaultmidpunct}
{\mcitedefaultendpunct}{\mcitedefaultseppunct}\relax
\EndOfBibitem
\bibitem[Cohen \latin{et~al.}(2008)Cohen, Mori-S{\'a}nchez, and
  Yang]{Cohen2008}
Cohen,~A.~J.; Mori-S{\'a}nchez,~P.; Yang,~W. Insights into Current Limitations
  of Density Functional Theory. \emph{Science} \textbf{2008}, \emph{321},
  792--794\relax
\mciteBstWouldAddEndPuncttrue
\mciteSetBstMidEndSepPunct{\mcitedefaultmidpunct}
{\mcitedefaultendpunct}{\mcitedefaultseppunct}\relax
\EndOfBibitem
\bibitem[Mori-S\'anchez \latin{et~al.}(2009)Mori-S\'anchez, Cohen, and
  Yang]{MoriSanchez2009}
Mori-S\'anchez,~P.; Cohen,~A.~J.; Yang,~W. Discontinuous Nature of the
  Exchange-Correlation Functional in Strongly Correlated Systems. \emph{Phys.
  Rev. Lett.} \textbf{2009}, \emph{102}, 066403\relax
\mciteBstWouldAddEndPuncttrue
\mciteSetBstMidEndSepPunct{\mcitedefaultmidpunct}
{\mcitedefaultendpunct}{\mcitedefaultseppunct}\relax
\EndOfBibitem
\bibitem[Kohn and Sham(1965)Kohn, and Sham]{KohnSham}
Kohn,~W.; Sham,~L.~J. Self-Consistent Equations Including Exchange and
  Correlation Effects. \emph{Phys. Rev.} \textbf{1965}, \emph{140},
  A1133--A1138\relax
\mciteBstWouldAddEndPuncttrue
\mciteSetBstMidEndSepPunct{\mcitedefaultmidpunct}
{\mcitedefaultendpunct}{\mcitedefaultseppunct}\relax
\EndOfBibitem
\bibitem[Perdew \latin{et~al.}(1996)Perdew, Burke, and Ernzerhof]{GGA}
Perdew,~J.~P.; Burke,~K.; Ernzerhof,~M. Generalized gradient approximation made
  simple. \emph{Phys. Rev. Lett.} \textbf{1996}, \emph{77}, 3865--3868\relax
\mciteBstWouldAddEndPuncttrue
\mciteSetBstMidEndSepPunct{\mcitedefaultmidpunct}
{\mcitedefaultendpunct}{\mcitedefaultseppunct}\relax
\EndOfBibitem
\bibitem[Becke(1988)]{Becke1988}
Becke,~A.~D. Density-functional exchange-energy approximation with correct
  asymptotic behavior. \emph{Phys. Rev. A} \textbf{1988}, \emph{38},
  3098--3100\relax
\mciteBstWouldAddEndPuncttrue
\mciteSetBstMidEndSepPunct{\mcitedefaultmidpunct}
{\mcitedefaultendpunct}{\mcitedefaultseppunct}\relax
\EndOfBibitem
\bibitem[Bruneval(2009)]{Bruneval2009}
Bruneval,~F. {$GW$} Approximation of the Many-Body Problem and Changes in the
  Particle Number. \emph{Phys. Rev. Lett.} \textbf{2009}, \emph{103},
  176403\relax
\mciteBstWouldAddEndPuncttrue
\mciteSetBstMidEndSepPunct{\mcitedefaultmidpunct}
{\mcitedefaultendpunct}{\mcitedefaultseppunct}\relax
\EndOfBibitem
\bibitem[Hellgren \latin{et~al.}(2012)Hellgren, Rohr, and
  Gross]{Hellgren2012-2}
Hellgren,~M.; Rohr,~D.~R.; Gross,~E. K.~U. Correlation potentials for molecular
  bond dissociation within the self-consistent random phase approximation.
  \emph{J. Chem. Phys.} \textbf{2012}, \emph{136}, 034106\relax
\mciteBstWouldAddEndPuncttrue
\mciteSetBstMidEndSepPunct{\mcitedefaultmidpunct}
{\mcitedefaultendpunct}{\mcitedefaultseppunct}\relax
\EndOfBibitem
\bibitem[Caruso \latin{et~al.}(2013)Caruso, Rohr, Hellgren, Ren, Rinke, Rubio,
  and Scheffler]{Caruso2013}
Caruso,~F.; Rohr,~D.~R.; Hellgren,~M.; Ren,~X.; Rinke,~P.; Rubio,~A.;
  Scheffler,~M. Bond Breaking and Bond Formation: How Electron Correlation is
  Captured in Many-Body Perturbation Theory and Density-Functional Theory.
  \emph{Phys. Rev. Lett.} \textbf{2013}, \emph{110}, 146403\relax
\mciteBstWouldAddEndPuncttrue
\mciteSetBstMidEndSepPunct{\mcitedefaultmidpunct}
{\mcitedefaultendpunct}{\mcitedefaultseppunct}\relax
\EndOfBibitem
\bibitem[Colonna \latin{et~al.}(2016)Colonna, Hellgren, and
  de~Gironcoli]{Colonna2016}
Colonna,~N.; Hellgren,~M.; de~Gironcoli,~S. Molecular bonding with the RPAx:
  From weak dispersion forces to strong correlation. \emph{Phys. Rev. B}
  \textbf{2016}, \emph{93}, 195108\relax
\mciteBstWouldAddEndPuncttrue
\mciteSetBstMidEndSepPunct{\mcitedefaultmidpunct}
{\mcitedefaultendpunct}{\mcitedefaultseppunct}\relax
\EndOfBibitem
\bibitem[Perdew \latin{et~al.}(1982)Perdew, Parr, Levy, and Balduz]{Perdew1982}
Perdew,~J.~P.; Parr,~R.~G.; Levy,~M.; Balduz,~J.~L. Density-Functional Theory
  for Fractional Particle Number: Derivative Discontinuities of the Energy.
  \emph{Phys. Rev. Lett.} \textbf{1982}, \emph{49}, 1691--1694\relax
\mciteBstWouldAddEndPuncttrue
\mciteSetBstMidEndSepPunct{\mcitedefaultmidpunct}
{\mcitedefaultendpunct}{\mcitedefaultseppunct}\relax
\EndOfBibitem
\bibitem[Makmal \latin{et~al.}(2011)Makmal, K\"ummel, and Kronik]{Makmal2011}
Makmal,~A.; K\"ummel,~S.; Kronik,~L. Dissociation of diatomic molecules and the
  exact-exchange Kohn-Sham potential: The case of LiF. \emph{Phys. Rev. A}
  \textbf{2011}, \emph{83}, 062512\relax
\mciteBstWouldAddEndPuncttrue
\mciteSetBstMidEndSepPunct{\mcitedefaultmidpunct}
{\mcitedefaultendpunct}{\mcitedefaultseppunct}\relax
\EndOfBibitem
\bibitem[Ruzsinszky \latin{et~al.}(2006)Ruzsinszky, Perdew, Csonka, Vydrov, and
  Scuseria]{Ruzsinszky2006}
Ruzsinszky,~A.; Perdew,~J.~P.; Csonka,~G.~I.; Vydrov,~O.~A.; Scuseria,~G.~E.
  Spurious fractional charge on dissociated atoms: Pervasive and resilient
  self-interaction error of common density functionals. \emph{J. Chem. Phys.}
  \textbf{2006}, \emph{125}, 194112\relax
\mciteBstWouldAddEndPuncttrue
\mciteSetBstMidEndSepPunct{\mcitedefaultmidpunct}
{\mcitedefaultendpunct}{\mcitedefaultseppunct}\relax
\EndOfBibitem
\bibitem[Vydrov \latin{et~al.}(2007)Vydrov, Scuseria, and Perdew]{Vydrov2007}
Vydrov,~O.~A.; Scuseria,~G.~E.; Perdew,~J.~P. Tests of functionals for systems
  with fractional electron number. \emph{J. Chem. Phys.} \textbf{2007},
  \emph{126}, 154109\relax
\mciteBstWouldAddEndPuncttrue
\mciteSetBstMidEndSepPunct{\mcitedefaultmidpunct}
{\mcitedefaultendpunct}{\mcitedefaultseppunct}\relax
\EndOfBibitem
\bibitem[Cohen \latin{et~al.}(2007)Cohen, Mori-S\'{a}nchez, and
  Yang]{Cohen2007}
Cohen,~A.~J.; Mori-S\'{a}nchez,~P.; Yang,~W. Development of
  exchange-correlation functionals with minimal many-electron self-interaction
  error. \emph{J. Chem. Phys.} \textbf{2007}, \emph{126}, 191109\relax
\mciteBstWouldAddEndPuncttrue
\mciteSetBstMidEndSepPunct{\mcitedefaultmidpunct}
{\mcitedefaultendpunct}{\mcitedefaultseppunct}\relax
\EndOfBibitem
\bibitem[Baer \latin{et~al.}(2010)Baer, Livshits, and Salzner]{Baer2010}
Baer,~R.; Livshits,~E.; Salzner,~U. Tuned Range-Separated Hybrids in Density
  Functional Theory. \emph{Annu. Rev. Phys. Chem.} \textbf{2010}, \emph{61},
  85--109, {PMID:} 20055678\relax
\mciteBstWouldAddEndPuncttrue
\mciteSetBstMidEndSepPunct{\mcitedefaultmidpunct}
{\mcitedefaultendpunct}{\mcitedefaultseppunct}\relax
\EndOfBibitem
\bibitem[Tsuneda \latin{et~al.}(2010)Tsuneda, Song, Suzuki, and
  Hirao]{Tsuneda2010}
Tsuneda,~T.; Song,~J.-W.; Suzuki,~S.; Hirao,~K. On Koopmans' theorem in density
  functional theory. \emph{J. Chem. Phys.} \textbf{2010}, \emph{133},
  174101\relax
\mciteBstWouldAddEndPuncttrue
\mciteSetBstMidEndSepPunct{\mcitedefaultmidpunct}
{\mcitedefaultendpunct}{\mcitedefaultseppunct}\relax
\EndOfBibitem
\bibitem[Johnson and Contreras-Garc\'{\i}a(2011)Johnson, and
  Contreras-Garc\'{\i}a]{Johnson2011}
Johnson,~E.~R.; Contreras-Garc\'{\i}a,~J. Communication: A density functional
  with accurate fractional-charge and fractional-spin behaviour for
  s-electrons. \emph{J. Chem. Phys.} \textbf{2011}, \emph{135}, 081103\relax
\mciteBstWouldAddEndPuncttrue
\mciteSetBstMidEndSepPunct{\mcitedefaultmidpunct}
{\mcitedefaultendpunct}{\mcitedefaultseppunct}\relax
\EndOfBibitem
\bibitem[Kraisler and Kronik(2013)Kraisler, and Kronik]{Kraisler2013}
Kraisler,~E.; Kronik,~L. Piecewise Linearity of Approximate Density Functionals
  Revisited: Implications for Frontier Orbital Energies. \emph{Phys. Rev.
  Lett.} \textbf{2013}, \emph{110}, 126403\relax
\mciteBstWouldAddEndPuncttrue
\mciteSetBstMidEndSepPunct{\mcitedefaultmidpunct}
{\mcitedefaultendpunct}{\mcitedefaultseppunct}\relax
\EndOfBibitem
\bibitem[Steinmann and Yang(2013)Steinmann, and Yang]{Steinmann2013}
Steinmann,~S.~N.; Yang,~W. Wave function methods for fractional electrons.
  \emph{J. Chem. Phys.} \textbf{2013}, \emph{139}, 074107\relax
\mciteBstWouldAddEndPuncttrue
\mciteSetBstMidEndSepPunct{\mcitedefaultmidpunct}
{\mcitedefaultendpunct}{\mcitedefaultseppunct}\relax
\EndOfBibitem
\bibitem[Zheng \latin{et~al.}(2013)Zheng, Zhou, and Yang]{Zheng2013}
Zheng,~X.; Zhou,~T.; Yang,~W. A nonempirical scaling correction approach for
  density functional methods involving substantial amount of Hartree--Fock
  exchange. \emph{J. Chem. Phys.} \textbf{2013}, \emph{138}, 174105\relax
\mciteBstWouldAddEndPuncttrue
\mciteSetBstMidEndSepPunct{\mcitedefaultmidpunct}
{\mcitedefaultendpunct}{\mcitedefaultseppunct}\relax
\EndOfBibitem
\bibitem[Malet and Gori-Giorgi(2012)Malet, and Gori-Giorgi]{Malet2012}
Malet,~F.; Gori-Giorgi,~P. Strong Correlation in Kohn-Sham Density Functional
  Theory. \emph{Phys. Rev. Lett.} \textbf{2012}, \emph{109}, 246402\relax
\mciteBstWouldAddEndPuncttrue
\mciteSetBstMidEndSepPunct{\mcitedefaultmidpunct}
{\mcitedefaultendpunct}{\mcitedefaultseppunct}\relax
\EndOfBibitem
\bibitem[Vuckovic and Gori-Giorgi(2017)Vuckovic, and Gori-Giorgi]{Vuckovic2017}
Vuckovic,~S.; Gori-Giorgi,~P. Simple Fully Nonlocal Density Functionals for
  Electronic Repulsion Energy. \emph{J. Phys. Chem. Lett.} \textbf{2017},
  \emph{8}, 2799--2805, PMID: 28581751\relax
\mciteBstWouldAddEndPuncttrue
\mciteSetBstMidEndSepPunct{\mcitedefaultmidpunct}
{\mcitedefaultendpunct}{\mcitedefaultseppunct}\relax
\EndOfBibitem
\bibitem[L\"owdin and Shull(1956)L\"owdin, and Shull]{lowdinRDMFT}
L\"owdin,~P.-O.; Shull,~H. Natural Orbitals in the Quantum Theory of
  Two-Electron Systems. \emph{Phys. Rev.} \textbf{1956}, \emph{101},
  1730--1739\relax
\mciteBstWouldAddEndPuncttrue
\mciteSetBstMidEndSepPunct{\mcitedefaultmidpunct}
{\mcitedefaultendpunct}{\mcitedefaultseppunct}\relax
\EndOfBibitem
\bibitem[Donnelly and Parr(1978)Donnelly, and Parr]{Donnelly1978}
Donnelly,~R.~A.; Parr,~R.~G. Elementary properties of an energy functional of
  the first-order reduced density matrix. \emph{J. Chem. Phys.} \textbf{1978},
  \emph{69}, 4431--4439\relax
\mciteBstWouldAddEndPuncttrue
\mciteSetBstMidEndSepPunct{\mcitedefaultmidpunct}
{\mcitedefaultendpunct}{\mcitedefaultseppunct}\relax
\EndOfBibitem
\bibitem[Sharp and Horton(1953)Sharp, and Horton]{OEP1}
Sharp,~R.~T.; Horton,~G.~K. A Variational Approach to the Unipotential
  Many-Electron Problem. \emph{Phys. Rev.} \textbf{1953}, \emph{90},
  317--317\relax
\mciteBstWouldAddEndPuncttrue
\mciteSetBstMidEndSepPunct{\mcitedefaultmidpunct}
{\mcitedefaultendpunct}{\mcitedefaultseppunct}\relax
\EndOfBibitem
\bibitem[Talman and Shadwick(1976)Talman, and Shadwick]{OEP2}
Talman,~J.~D.; Shadwick,~W.~F. Optimized effective atomic central potential.
  \emph{Phys. Rev. A} \textbf{1976}, \emph{14}, 36--40\relax
\mciteBstWouldAddEndPuncttrue
\mciteSetBstMidEndSepPunct{\mcitedefaultmidpunct}
{\mcitedefaultendpunct}{\mcitedefaultseppunct}\relax
\EndOfBibitem
\bibitem[Sch\"onhammer and Gunnarsson(1987)Sch\"onhammer, and
  Gunnarsson]{Schoenhammer1987}
Sch\"onhammer,~K.; Gunnarsson,~O. Discontinuity of the exchange-correlation
  potential in density functional theory. \emph{J. Phys. C: Solid State Phys.}
  \textbf{1987}, \emph{20}, 3675\relax
\mciteBstWouldAddEndPuncttrue
\mciteSetBstMidEndSepPunct{\mcitedefaultmidpunct}
{\mcitedefaultendpunct}{\mcitedefaultseppunct}\relax
\EndOfBibitem
\bibitem[Krieger \latin{et~al.}(1992)Krieger, Li, and Iafrate]{KLI1992}
Krieger,~J.~B.; Li,~Y.; Iafrate,~G.~J. Construction and application of an
  accurate local spin-polarized Kohn-Sham potential with integer discontinuity:
  Exchange-only theory. \emph{Phys. Rev. A} \textbf{1992}, \emph{45},
  101--126\relax
\mciteBstWouldAddEndPuncttrue
\mciteSetBstMidEndSepPunct{\mcitedefaultmidpunct}
{\mcitedefaultendpunct}{\mcitedefaultseppunct}\relax
\EndOfBibitem
\bibitem[Gori-Giorgi and Savin(2009)Gori-Giorgi, and Savin]{GoriGiorgi2009}
Gori-Giorgi,~P.; Savin,~A. Study of the discontinuity of the
  exchange-correlation potential in an exactly soluble case. \emph{Int. J.
  Quantum Chem.} \textbf{2009}, \emph{109}, 2410--2415\relax
\mciteBstWouldAddEndPuncttrue
\mciteSetBstMidEndSepPunct{\mcitedefaultmidpunct}
{\mcitedefaultendpunct}{\mcitedefaultseppunct}\relax
\EndOfBibitem
\bibitem[Hellgren and Gross(2012)Hellgren, and Gross]{Hellgren2012}
Hellgren,~M.; Gross,~E. K.~U. Discontinuities of the exchange-correlation
  kernel and charge-transfer excitations in time-dependent density-functional
  theory. \emph{Phys. Rev. A} \textbf{2012}, \emph{85}, 022514\relax
\mciteBstWouldAddEndPuncttrue
\mciteSetBstMidEndSepPunct{\mcitedefaultmidpunct}
{\mcitedefaultendpunct}{\mcitedefaultseppunct}\relax
\EndOfBibitem
\bibitem[Hellgren and Gross(2013)Hellgren, and Gross]{Hellgren2013}
Hellgren,~M.; Gross,~E. K.~U. Discontinuous functional for linear-response
  time-dependent density-functional theory: The exact-exchange kernel and
  approximate forms. \emph{Phys. Rev. A} \textbf{2013}, \emph{88}, 052507\relax
\mciteBstWouldAddEndPuncttrue
\mciteSetBstMidEndSepPunct{\mcitedefaultmidpunct}
{\mcitedefaultendpunct}{\mcitedefaultseppunct}\relax
\EndOfBibitem
\bibitem[Gidopoulos \latin{et~al.}(2002)Gidopoulos, Papaconstantinou, and
  Gross]{Gidopoulos2002}
Gidopoulos,~N.~I.; Papaconstantinou,~P.~G.; Gross,~E. K.~U. Spurious
  Interactions, and Their Correction, in the Ensemble-Kohn-Sham Scheme for
  Excited States. \emph{Phys. Rev. Lett.} \textbf{2002}, \emph{88},
  033003\relax
\mciteBstWouldAddEndPuncttrue
\mciteSetBstMidEndSepPunct{\mcitedefaultmidpunct}
{\mcitedefaultendpunct}{\mcitedefaultseppunct}\relax
\EndOfBibitem
\bibitem[Grabowski \latin{et~al.}(2002)Grabowski, Hirata, Ivanov, and
  Bartlett]{Grabowski2002}
Grabowski,~I.; Hirata,~S.; Ivanov,~S.; Bartlett,~R.~J. Ab initio density
  functional theory: OEP-MBPT(2). A new orbital-dependent correlation
  functional. \emph{J. Chem. Phys.} \textbf{2002}, \emph{116}, 4415--4425\relax
\mciteBstWouldAddEndPuncttrue
\mciteSetBstMidEndSepPunct{\mcitedefaultmidpunct}
{\mcitedefaultendpunct}{\mcitedefaultseppunct}\relax
\EndOfBibitem
\bibitem[Lathiotakis \latin{et~al.}(2014)Lathiotakis, Helbig, Rubio, and
  Gidopoulos]{lRDMFT}
Lathiotakis,~N.~N.; Helbig,~N.; Rubio,~A.; Gidopoulos,~N.~I. Local
  reduced-density-matrix-functional theory: Incorporating static correlation
  effects in Kohn-Sham equations. \emph{Phys. Rev. A} \textbf{2014}, \emph{90},
  032511\relax
\mciteBstWouldAddEndPuncttrue
\mciteSetBstMidEndSepPunct{\mcitedefaultmidpunct}
{\mcitedefaultendpunct}{\mcitedefaultseppunct}\relax
\EndOfBibitem
\bibitem[Muller and Desjarlais(2006)Muller, and Desjarlais]{MCOEP1}
Muller,~R.~P.; Desjarlais,~M.~P. Optimized effective potential from a
  correlated wave function: Optimized effective potential-generalized valence
  bond (OEP-GVB). \emph{J. Chem. Phys.} \textbf{2006}, \emph{125}, 054101\relax
\mciteBstWouldAddEndPuncttrue
\mciteSetBstMidEndSepPunct{\mcitedefaultmidpunct}
{\mcitedefaultendpunct}{\mcitedefaultseppunct}\relax
\EndOfBibitem
\bibitem[Weimer \latin{et~al.}(2008)Weimer, Della~Sala, and G\"orling]{MCOEP2}
Weimer,~M.; Della~Sala,~F.; G\"orling,~A. Multiconfiguration optimized
  effective potential method for a density-functional treatment of static
  correlation. \emph{J. Chem. Phys.} \textbf{2008}, \emph{128}, 144109\relax
\mciteBstWouldAddEndPuncttrue
\mciteSetBstMidEndSepPunct{\mcitedefaultmidpunct}
{\mcitedefaultendpunct}{\mcitedefaultseppunct}\relax
\EndOfBibitem
\bibitem[Goddard~III \latin{et~al.}(1973)Goddard~III, Dunning~Jr, Hunt, and
  Hay]{GVB}
Goddard~III,~W.~A.; Dunning~Jr,~T.~H.; Hunt,~W.~J.; Hay,~P.~J. Generalized
  valence bond description of bonding in low-lying states of molecules.
  \emph{Acc. Chem. Res.} \textbf{1973}, \emph{6}, 368--376\relax
\mciteBstWouldAddEndPuncttrue
\mciteSetBstMidEndSepPunct{\mcitedefaultmidpunct}
{\mcitedefaultendpunct}{\mcitedefaultseppunct}\relax
\EndOfBibitem
\bibitem[Hellgren and von Barth(2007)Hellgren, and von Barth]{hellgren07}
Hellgren,~M.; von Barth,~U. Correlation potential in density functional theory
  at the GWA level: Spherical atoms. \emph{Phys. Rev. B} \textbf{2007},
  \emph{76}, 075107\relax
\mciteBstWouldAddEndPuncttrue
\mciteSetBstMidEndSepPunct{\mcitedefaultmidpunct}
{\mcitedefaultendpunct}{\mcitedefaultseppunct}\relax
\EndOfBibitem
\bibitem[Hellgren and von Barth(2009)Hellgren, and von Barth]{hellgren09}
Hellgren,~M.; von Barth,~U. Exact-exchange kernel of time-dependent density
  functional theory: Frequency dependence and photoabsorption spectra of atoms.
  \emph{J. Chem. Phys.} \textbf{2009}, \emph{131}, 044110\relax
\mciteBstWouldAddEndPuncttrue
\mciteSetBstMidEndSepPunct{\mcitedefaultmidpunct}
{\mcitedefaultendpunct}{\mcitedefaultseppunct}\relax
\EndOfBibitem
\bibitem[Gross \latin{et~al.}(1988)Gross, Oliveira, and Kohn]{GOK1}
Gross,~E. K.~U.; Oliveira,~L.~N.; Kohn,~W. {Rayleigh}-Ritz variational
  principle for ensembles of fractionally occupied states. \emph{Phys. Rev. A}
  \textbf{1988}, \emph{37}, 2805--2808\relax
\mciteBstWouldAddEndPuncttrue
\mciteSetBstMidEndSepPunct{\mcitedefaultmidpunct}
{\mcitedefaultendpunct}{\mcitedefaultseppunct}\relax
\EndOfBibitem
\bibitem[Gross \latin{et~al.}(1988)Gross, Oliveira, and Kohn]{GOK2}
Gross,~E. K.~U.; Oliveira,~L.~N.; Kohn,~W. Density-functional theory for
  ensembles of fractionally occupied states. I. Basic formalism. \emph{Phys.
  Rev. A} \textbf{1988}, \emph{37}, 2809--2820\relax
\mciteBstWouldAddEndPuncttrue
\mciteSetBstMidEndSepPunct{\mcitedefaultmidpunct}
{\mcitedefaultendpunct}{\mcitedefaultseppunct}\relax
\EndOfBibitem
\bibitem[Gould and Pittalis(2017)Gould, and Pittalis]{Gould2017-Limits}
Gould,~T.; Pittalis,~S. Hartree and Exchange in Ensemble Density Functional
  Theory: Avoiding the Nonuniqueness Disaster. \emph{Phys. Rev. Lett.}
  \textbf{2017}, \emph{119}, 243001\relax
\mciteBstWouldAddEndPuncttrue
\mciteSetBstMidEndSepPunct{\mcitedefaultmidpunct}
{\mcitedefaultendpunct}{\mcitedefaultseppunct}\relax
\EndOfBibitem
\bibitem[Gould and Pittalis(2019)Gould, and Pittalis]{Gould2019-DD}
Gould,~T.; Pittalis,~S. Density-Driven Correlations in Many-Electron Ensembles:
  Theory and Application for Excited States. \emph{Phys. Rev. Lett.}
  \textbf{2019}, \emph{123}, 016401\relax
\mciteBstWouldAddEndPuncttrue
\mciteSetBstMidEndSepPunct{\mcitedefaultmidpunct}
{\mcitedefaultendpunct}{\mcitedefaultseppunct}\relax
\EndOfBibitem
\bibitem[Buijse and Baerends(2002)Buijse, and Baerends]{bb}
Buijse,~M.~A.; Baerends,~E.~J. An approximate exchange-correlation hole density
  as a functional of the natural orbitals. \emph{Mol. Phys.} \textbf{2002},
  \emph{100}, 401\relax
\mciteBstWouldAddEndPuncttrue
\mciteSetBstMidEndSepPunct{\mcitedefaultmidpunct}
{\mcitedefaultendpunct}{\mcitedefaultseppunct}\relax
\EndOfBibitem
\bibitem[M\"uller(1984)]{muller}
M\"uller,~A. M.~K. Explicit Approximate Relation between Reduced Two- and
  One-Particle Density Matrices. \emph{Phys. Lett. A} \textbf{1984},
  \emph{105}, 446\relax
\mciteBstWouldAddEndPuncttrue
\mciteSetBstMidEndSepPunct{\mcitedefaultmidpunct}
{\mcitedefaultendpunct}{\mcitedefaultseppunct}\relax
\EndOfBibitem
\bibitem[Kamil \latin{et~al.}(2016)Kamil, Schade, Pruschke, and
  Bl\"ochl]{hubbardRDMFT}
Kamil,~E.; Schade,~R.; Pruschke,~T.; Bl\"ochl,~P.~E. Reduced density-matrix
  functionals applied to the Hubbard dimer. \emph{Phys. Rev. B} \textbf{2016},
  \emph{93}, 085141\relax
\mciteBstWouldAddEndPuncttrue
\mciteSetBstMidEndSepPunct{\mcitedefaultmidpunct}
{\mcitedefaultendpunct}{\mcitedefaultseppunct}\relax
\EndOfBibitem
\bibitem[Lathiotakis \latin{et~al.}(2009)Lathiotakis, Sharma, Dewhurst, Eich,
  Marques, and Gross]{powerRDMFT}
Lathiotakis,~N.~N.; Sharma,~S.; Dewhurst,~J.~K.; Eich,~F.~G.; Marques,~M.
  A.~L.; Gross,~E. K.~U. Density-matrix-power functional: Performance for
  finite systems and the homogeneous electron gas. \emph{Phys. Rev. A}
  \textbf{2009}, \emph{79}, 040501\relax
\mciteBstWouldAddEndPuncttrue
\mciteSetBstMidEndSepPunct{\mcitedefaultmidpunct}
{\mcitedefaultendpunct}{\mcitedefaultseppunct}\relax
\EndOfBibitem
\bibitem[Tempel \latin{et~al.}(2009)Tempel, Mart\'ines, and Maitra]{Maitra2009}
Tempel,~D.~G.; Mart\'ines,~T.~J.; Maitra,~N.~T. Revisiting Molecular
  Dissociation in Density FunctionalTheory: A Simple Model. \emph{J. Chem.
  Theory Comput.} \textbf{2009}, \emph{5}, 770--780\relax
\mciteBstWouldAddEndPuncttrue
\mciteSetBstMidEndSepPunct{\mcitedefaultmidpunct}
{\mcitedefaultendpunct}{\mcitedefaultseppunct}\relax
\EndOfBibitem
\bibitem[Helbig \latin{et~al.}(2011)Helbig, Fuks, Casula, Verstraete, Marques,
  Tokatly, and Rubio]{helbig_soft}
Helbig,~N.; Fuks,~J.~I.; Casula,~M.; Verstraete,~M.~J.; Marques,~M. A.~L.;
  Tokatly,~I.~V.; Rubio,~A. Density functional theory beyond the linear regime:
  Validating an adiabatic local density approximation. \emph{Phys. Rev. A}
  \textbf{2011}, \emph{83}, 032503\relax
\mciteBstWouldAddEndPuncttrue
\mciteSetBstMidEndSepPunct{\mcitedefaultmidpunct}
{\mcitedefaultendpunct}{\mcitedefaultseppunct}\relax
\EndOfBibitem
\bibitem[Oueis and Wasserman(2018)Oueis, and Wasserman]{Oueis2018}
Oueis,~Y.; Wasserman,~A. Exact partition potential for model systems of
  interacting electrons in 1-D. \emph{Euro. Phys. J. B} \textbf{2018},
  \emph{91}, 247\relax
\mciteBstWouldAddEndPuncttrue
\mciteSetBstMidEndSepPunct{\mcitedefaultmidpunct}
{\mcitedefaultendpunct}{\mcitedefaultseppunct}\relax
\EndOfBibitem
\bibitem[Marques \latin{et~al.}(2006)Marques, Castro, Bertsch, and
  Rubio]{octopus}
Marques,~M. A.~L.; Castro,~A.; Bertsch,~G.~F.; Rubio,~A. {\em octopus}: a tool
  for the application of time-dependent density functional theory. \emph{Phys.
  Status Solidi B} \textbf{2006}, \emph{243}, 2465\relax
\mciteBstWouldAddEndPuncttrue
\mciteSetBstMidEndSepPunct{\mcitedefaultmidpunct}
{\mcitedefaultendpunct}{\mcitedefaultseppunct}\relax
\EndOfBibitem
\bibitem[Sham and Schl\"uter(1983)Sham, and Schl\"uter]{lss}
Sham,~L.~J.; Schl\"uter,~M. Density-Functional Theory of the Energy Gap.
  \emph{Phys. Rev. Lett.} \textbf{1983}, \emph{51}, 1888--1891\relax
\mciteBstWouldAddEndPuncttrue
\mciteSetBstMidEndSepPunct{\mcitedefaultmidpunct}
{\mcitedefaultendpunct}{\mcitedefaultseppunct}\relax
\EndOfBibitem
\bibitem[Casida(1995)]{casida}
Casida,~M.~E. Generalization of the optimized-effective-potential model to
  include electron correlation: A variational derivation of the Sham-Schl\"uter
  equation for the exact exchange-correlation potential. \emph{Phys. Rev. A}
  \textbf{1995}, \emph{51}, 2005--2013\relax
\mciteBstWouldAddEndPuncttrue
\mciteSetBstMidEndSepPunct{\mcitedefaultmidpunct}
{\mcitedefaultendpunct}{\mcitedefaultseppunct}\relax
\EndOfBibitem
\bibitem[von Barth \latin{et~al.}(2005)von Barth, Dahlen, van Leeuwen, and
  Stefanucci]{vonBarth05}
von Barth,~U.; Dahlen,~N.~E.; van Leeuwen,~R.; Stefanucci,~G. Conserving
  approximations in time-dependent density functional theory. \emph{Phys. Rev.
  B} \textbf{2005}, \emph{72}, 235109\relax
\mciteBstWouldAddEndPuncttrue
\mciteSetBstMidEndSepPunct{\mcitedefaultmidpunct}
{\mcitedefaultendpunct}{\mcitedefaultseppunct}\relax
\EndOfBibitem
\bibitem[Maitra(2005)]{maitraCT}
Maitra,~N.~T. Undoing static correlation: Long-range charge transfer in
  time-dependent density-functional theory. \emph{J. Chem. Phys.}
  \textbf{2005}, \emph{122}, 234104\relax
\mciteBstWouldAddEndPuncttrue
\mciteSetBstMidEndSepPunct{\mcitedefaultmidpunct}
{\mcitedefaultendpunct}{\mcitedefaultseppunct}\relax
\EndOfBibitem
\bibitem[Hait and Head-Gordon(2018)Hait, and Head-Gordon]{Hait2018}
Hait,~D.; Head-Gordon,~M. How accurate is density functional theory at
  predicting dipole moments? an assessment using a new database of 200
  benchmark values. \emph{Journal of chemical theory and computation}
  \textbf{2018}, \emph{14}, 1969--1981\relax
\mciteBstWouldAddEndPuncttrue
\mciteSetBstMidEndSepPunct{\mcitedefaultmidpunct}
{\mcitedefaultendpunct}{\mcitedefaultseppunct}\relax
\EndOfBibitem
\bibitem[Mori-S\'anchez \latin{et~al.}(2012)Mori-S\'anchez, Cohen, and
  Yang]{MORI-RPA}
Mori-S\'anchez,~P.; Cohen,~A.~J.; Yang,~W. Failure of the
  random-phase-approximation correlation energy. \emph{Phys. Rev. A}
  \textbf{2012}, \emph{85}, 042507\relax
\mciteBstWouldAddEndPuncttrue
\mciteSetBstMidEndSepPunct{\mcitedefaultmidpunct}
{\mcitedefaultendpunct}{\mcitedefaultseppunct}\relax
\EndOfBibitem
\bibitem[Buijse \latin{et~al.}(1989)Buijse, Baerends, and Snijders]{peak1}
Buijse,~M.~A.; Baerends,~E.~J.; Snijders,~J.~G. Analysis of correlation in
  terms of exact local potentials: Applications to two-electron systems.
  \emph{Phys. Rev. A} \textbf{1989}, \emph{40}, 4190--4202\relax
\mciteBstWouldAddEndPuncttrue
\mciteSetBstMidEndSepPunct{\mcitedefaultmidpunct}
{\mcitedefaultendpunct}{\mcitedefaultseppunct}\relax
\EndOfBibitem
\bibitem[Gritsenko and Baerends(1996)Gritsenko, and Baerends]{peak2}
Gritsenko,~O.~V.; Baerends,~E.~J. Effect of molecular dissociation on the
  exchange-correlation Kohn-Sham potential. \emph{Phys. Rev. A} \textbf{1996},
  \emph{54}, 1957--1972\relax
\mciteBstWouldAddEndPuncttrue
\mciteSetBstMidEndSepPunct{\mcitedefaultmidpunct}
{\mcitedefaultendpunct}{\mcitedefaultseppunct}\relax
\EndOfBibitem
\bibitem[van Gisbergen \latin{et~al.}(1999)van Gisbergen, Schipper, Gritsenko,
  Baerends, Snijders, Champagne, and Kirtman]{gisbergen99}
van Gisbergen,~S. J.~A.; Schipper,~P. R.~T.; Gritsenko,~O.~V.; Baerends,~E.~J.;
  Snijders,~J.~G.; Champagne,~B.; Kirtman,~B. Electric Field Dependence of the
  Exchange-Correlation Potential in Molecular Chains. \emph{Phys. Rev. Lett.}
  \textbf{1999}, \emph{83}, 694--697\relax
\mciteBstWouldAddEndPuncttrue
\mciteSetBstMidEndSepPunct{\mcitedefaultmidpunct}
{\mcitedefaultendpunct}{\mcitedefaultseppunct}\relax
\EndOfBibitem
\bibitem[K\"ummel \latin{et~al.}(2004)K\"ummel, Kronik, and Perdew]{kummel04}
K\"ummel,~S.; Kronik,~L.; Perdew,~J.~P. Electrical Response of Molecular Chains
  from Density Functional Theory. \emph{Phys. Rev. Lett.} \textbf{2004},
  \emph{93}, 213002\relax
\mciteBstWouldAddEndPuncttrue
\mciteSetBstMidEndSepPunct{\mcitedefaultmidpunct}
{\mcitedefaultendpunct}{\mcitedefaultseppunct}\relax
\EndOfBibitem
\bibitem[Hodgson \latin{et~al.}(2017)Hodgson, Kraisler, Schild, and
  Gross]{stepshodgson}
Hodgson,~M. J.~P.; Kraisler,~E.; Schild,~A.; Gross,~E. K.~U. How Interatomic
  Steps in the Exact Kohn-Sham Potential Relate to Derivative Discontinuities
  of the Energy. \emph{J. Phys. Chem. Lett.} \textbf{2017}, \emph{8},
  5974\relax
\mciteBstWouldAddEndPuncttrue
\mciteSetBstMidEndSepPunct{\mcitedefaultmidpunct}
{\mcitedefaultendpunct}{\mcitedefaultseppunct}\relax
\EndOfBibitem
\bibitem[Cohen and Mori-S\'anchez(2014)Cohen, and Mori-S\'anchez]{fracnuclei}
Cohen,~A.~J.; Mori-S\'anchez,~P. Dramatic changes in electronic structure
  revealed by fractionally charged nuclei. \emph{J. Chem. Phys.} \textbf{2014},
  \emph{140}, 044110\relax
\mciteBstWouldAddEndPuncttrue
\mciteSetBstMidEndSepPunct{\mcitedefaultmidpunct}
{\mcitedefaultendpunct}{\mcitedefaultseppunct}\relax
\EndOfBibitem
\bibitem[Gould \latin{et~al.}(2018)Gould, Kronik, and Pittalis]{Gould2018-CT}
Gould,~T.; Kronik,~L.; Pittalis,~S. Charge transfer excitations from exact and
  approximate ensemble Kohn-Sham theory. \emph{J. Chem. Phys.} \textbf{2018},
  \emph{148}, 174101\relax
\mciteBstWouldAddEndPuncttrue
\mciteSetBstMidEndSepPunct{\mcitedefaultmidpunct}
{\mcitedefaultendpunct}{\mcitedefaultseppunct}\relax
\EndOfBibitem
\bibitem[Kim \latin{et~al.}(2013)Kim, Sim, and Burke]{Kim2013}
Kim,~M.-C.; Sim,~E.; Burke,~K. Understanding and Reducing Errors in Density
  Functional Calculations. \emph{Phys. Rev. Lett.} \textbf{2013}, \emph{111},
  073003\relax
\mciteBstWouldAddEndPuncttrue
\mciteSetBstMidEndSepPunct{\mcitedefaultmidpunct}
{\mcitedefaultendpunct}{\mcitedefaultseppunct}\relax
\EndOfBibitem
\end{mcitethebibliography}
\providecommand{\latin}[1]{#1}
\providecommand*\mcitethebibliography{\thebibliography}
\csname @ifundefined\endcsname{endmcitethebibliography}
  {\let\endmcitethebibliography\endthebibliography}{}

\end{document}